 \documentstyle[epsf,twocolumn,prd,aps,floats]{revtex}

\begin{document}

\draft
\date{June 29, 2000}
%
%
\newcommand{\nc}{\newcommand}
\nc{\bea}{\begin{eqnarray}}
\nc{\eea}{\end{eqnarray}}
\nc{\beq}{\begin{equation}}
\nc{\eeq}{\end{equation}}
\nc{\bi}{\begin{itemize}}
\nc{\ei}{\end{itemize}}
\nc{\la}[1]{\label{#1}}
\nc{\half}{\frac{1}{2}}
\nc{\fsky}{f_{\rm sky}}
\nc{\fwhm}{\theta_{\rm fwhm}}
\nc{\fwhmc}{\theta_{{\rm fwhm},c}}
\nc{\nad}{n_{\rm ad}}
\nc{\niso}{n_{\rm iso}}
\nc{\R}{{\cal{R}}}
\nc{\GeV}{\mbox{ GeV}}
\nc{\MeV}{\mbox{ MeV}}
\nc{\keV}{\mbox{ keV}}
\nc{\etal}{{\it et al.}}
\nc{\x}[1]{}
%
%

\nc{\AJ}[3]{{Astron.~J.\ }{{\bf #1}{, #2}{ (#3)}}}
\nc{\anap}[3]{{Astron.\ Astrophys.\ }{{\bf #1}{, #2}{ (#3)}}}
\nc{\ApJ}[3]{{Astrophys.~J.\ }{{\bf #1}{, #2}{ (#3)}}}
\nc{\apjl}[3]{{Astrophys.~J.\ Lett.\ }{{\bf #1}{, #2}{ (#3)}}}
\nc{\app}[3]{{Astropart.\ Phys.\ }{{\bf #1}{, #2}{ (#3)}}}
\nc{\araa}[3]{{Ann.\ Rev.\ Astron.\ Astrophys.\ }{{\bf #1}{, #2}{ (#3)}}}
\nc{\arns}[3]{{Ann.\ Rev.\ Nucl.\ Sci.\ }{{\bf #1}{, #2}{ (#3)}}}
\nc{\arnps}[3]{{Ann.\ Rev.\ Nucl.\ and Part.\ Sci.\ }{{\bf #1}{, #2}{ (#3)}}}
\nc{\epj}[3]{{Eur.\ Phys.\ J.\ }{{\bf #1}{, #2}{ (#3)}}}
\nc{\MNRAS}[3]{{Mon.\ Not.\ R.\ Astron.\ Soc.\ }{{\bf #1}{, #2}{ (#3)}}}
\nc{\mpl}[3]{{Mod.\ Phys.\ Lett.\ }{{\bf #1}{, #2}{ (#3)}}}
\nc{\Nat}[3]{{Nature }{{\bf #1}{, #2}{ (#3)}}}
\nc{\ncim}[3]{{Nuov.\ Cim.\ }{{\bf #1}{, #2}{ (#3)}}}
\nc{\nast}[3]{{New Astronomy }{{\bf #1}{, #2}{ (#3)}}}
\nc{\np}[3]{{Nucl.\ Phys.\ }{{\bf #1}{, #2}{ (#3)}}}
\nc{\pr}[3]{{Phys.\ Rev.\ }{{\bf #1}{, #2}{ (#3)}}}
\nc{\PRD}[3]{{Phys.\ Rev.\ D\ }{{\bf #1}{, #2}{ (#3)}}}
\nc{\PRL}[3]{{Phys.\ Rev.\ Lett.\ }{{\bf #1}{, #2}{ (#3)}}}
\nc{\PL}[3]{{Phys.\ Lett.\ }{{\bf #1}{, #2}{ (#3)}}}
\nc{\prep}[3]{{Phys.\ Rep.\ }{{\bf #1}{, #2}{ (#3)}}}
\nc{\RMP}[3]{{Rev.\ Mod.\ Phys.\ }{{\bf #1}{, #2}{ (#3)}}}
\nc{\rpp}[3]{{Rep.\ Prog.\ Phys.\ }{{\bf #1}{, #2}{ (#3)}}}
\nc{\ibid}[3]{{\it ibid.\ }{{\bf #1}{, #2}{ (#3)}}}

\wideabs{
\title{Limits on isocurvature fluctuations from Boomerang and MAXIMA
}

\author{Kari Enqvist\cite{mailk}}
\address{Department of Physics, University of Helsinki, and Helsinki Institute of Physics,\\
         P.O.Box 9, FIN-00014 University of Helsinki, Finland}

\author{Hannu Kurki-Suonio\cite{mailh}}
\address{Helsinki Institute of Physics,
         P.O.Box 9, FIN-00014 University of Helsinki, Finland}

\author{Jussi V\"{a}liviita\cite{mailv}}
\address{Department of Physics, University of Helsinki,
         P.O.Box 9, FIN-00014 University of Helsinki, Finland}

\maketitle

\begin{abstract}
We present the constraints on isocurvature fluctuations for a flat
universe implied by the Boomerang and Maxima-1
data on the anisotropy of the cosmic
microwave background. Because the new data defines the shape of
the angular power spectrum in the region of the first acoustic
peaks much more clearly than earlier data, even a tilted pure
isocurvature model is now ruled out. However, a mixed model with a
sizable isocurvature contribution remains allowed. We consider
primordial fluctuations with
different spectral indices for the adiabatic and isocurvature
perturbations, and find that the 95\% C.L. upper limit to the
isocurvature contribution to the low multipoles is $\alpha \leq
0.63$.  The upper limit to the contribution in the $\l \sim 200$
region is $\alpha_{200} \leq 0.13$.

\end{abstract}

\pacs{PACS numbers: 98.70.Vc, 98.80.Cq}

\vspace*{-10cm}
\noindent
\hspace*{15cm} \mbox{HIP-2000-30/TH}
\vspace*{9.5cm}

}

%
%
\section{Introduction}

Two recent balloon-borne experiments, Boomerang\cite{boom} and
Maxima-1\cite{max1}, have lent considerable support to the
inflationary paradigm. The Boomerang experiment, which was flown
over Antarctica for 10 days in 1999 and which measured the
temperature fluctuations over 1\%\ of the microwave sky, was the
first to provide firm evidence for the first (and possibly second)
acoustic peak. The angular power spectrum, obtained from a
preliminary analysis of the Boomerang data,  is compatible with
approximate scale invariance predicted by cosmic inflation. The
first acoustic peak is at the multipole $l\simeq 200$, implying a
flat universe with $\Omega \equiv \Omega_m+\Omega_\Lambda\simeq
1$ in the case of adiabatic fluctuations \cite{boom}.
The Boomerang results have been confirmed by
Maxima-1\cite{max1,max2}, which observed a 124 square degree patch
of the sky. Boomerang and Maxima-1 are the first precision
measurements of the cosmic microwave background (CMB) temperature
fluctuations at multipoles $l \gtrsim 30$. They will be followed by
the second generation satellite experiments Microwave Anisotropy Probe
(MAP) \cite{MAP} and
Planck\cite{Planck}.

The simplest one-field inflation model
typically predicts a
near scale invariant power spectrum with Gaussian fluctuations.
The perturbations generated by quantum fluctuations are
adiabatic, with the number density proportional to entropy density
so that $\delta (n/s) = 0$.  This is so because the
quantum fluctuations of the inflaton field are directly reflected
in perturbations of
the inflaton energy density.

More generally, in addition to adiabatic fluctuations, there is
also another fundamental fluctuation mode. It does not perturb the
total energy density on comoving hypersurfaces (orthogonal to
fluid flow), so that $\delta\rho=0$.  Since there is then no
curvature perturbation of the comoving hypersurfaces, it is called
an isocurvature fluctuation\cite{EB86,general}.  Such fluctuations
can arise in particle physics models where, during inflation, there
appear also other low mass fields in addition to the inflaton. For
instance, for a complex scalar field with a U(1) symmetry the
fluctuations of the phase do not affect the energy density but
would show up as perturbations in the conserved charge associated
with the symmetry. This is effectively the case in the minimally
supersymmetric model, and its extensions, which have several flat
directions in the potential by virtue of gauge invariance and
supersymmetry. The complex fields along these flat directions,
called Affleck-Dine fields\cite{ad}, will be subject to quantum
fluctuations similar to the inflaton. As the Affleck-Dine fields
typically carry global charges such as baryon number, in addition
to adiabatic perturbations, there will also be isocurvature
perturbations\cite{johncmb}. The Affleck-Dine condensate itself is
not stable but fragments into non-topological solitons\cite{ks2},
or Q-balls. These may be absolutely stable, in which case there
will be isocurvature fluctuations in the baryon number, or
unstable but long-lived, in which case isocurvature fluctuation
can be imprinted also on the cold dark matter (CDM)
particles\cite{bbb2}. Other particle physics motivated sources for
isocurvature fluctuations are axion models \cite{axion} or models
of inflation with more than one field\cite{infla}.

For isocurvature perturbations the overdensities  in a given
particle species are balanced by perturbations in other particle
species, such as radiation. At the last scattering surface (LSS)
the compensation for the isocurvature perturbations can be
maintained only for scales larger than the horizon, effectively
generating extra power to photon perturbations at small multipoles
$l$. As a consequence, the CMB angular power spectrum, $C_l$, of
isocurvature perturbations differs a great deal from adiabatic
perturbations, and a purely isocurvature CDM perturbation with a
scale free spectrum is clearly ruled out \cite{Stompor}
on the basis of the Cosmic Background Explorer (COBE)
normalization\cite{COBEdata,BunnWhite} and
$\sigma_8$, the observed amplitude of the rms mass fluctuations in
an $8h^{-1}$ Mpc sphere.

To match $\sigma_8$ with the COBE normalization requires a large
``blue" tilt $\niso \approx 2.2$.  Such a tilted isocurvature
model can be obtained from a reasonable inflation
model\cite{Peebles1} and agrees with most observational data. (The
main difficulty before Boomerang was that it has more tilt at low
multipoles than what COBE saw.) Thus it could be argued that a
pure isocurvature model was still acceptable\cite{Peebles2}.

A mixed perturbation, with a small isocurvature component is certainly
allowed and indeed motivated by particle physics, although the
forthcoming satellite experiments are expected to be able to
constrain the isocurvature amplitude with a high
precision\cite{eks,gb}.

The purpose of the present paper is to find out the constraints on
the isocurvature component as implied by the Boomerang and
Maxima-1 data. The isocurvature perturbation is highly degenerate
with the tensor perturbation, and for their separation one
probably has to wait for the polarization data from Planck\cite{eks}.
Here we will make the simplifying assumption that the tensor
contribution is negligible. This will be the case in so-called
small-field inflation models\cite{DKK97,LyRi99}. For large-field
inflation, the tensor/scalar ratio is $r \equiv C_2^T/C_2^{\rm ad}
\lesssim 7(1-\nad)$, so that the tensor contribution disappears
for a scale-free spectrum, $\nad \approx 1$. As we shall see, such
an assumption is consistent with the Boomerang and Maxima-1 data.

An isocurvature contribution is also somewhat degenerate with
reionization.  As we are looking for the upper limit to an isocurvature
contribution, we make the simplifying assumption that
reionization has a negligible effect.

\section{Adiabatic vs isocurvature perturbations}

Adiabatic perturbations\cite{MFB92,LL93} are
characterized by the quantity $\R$, which is related to the
curvature perturbation of the comoving hypersurface,
\beq
   \R_k = \frac{1}{4}\biggl(\frac{a}{k}\biggr)^2R_k^{(3)},
\eeq
where $a$ is the scale factor, $k$ is the comoving wave
number, and $R^{(3)}$ is the scalar curvature of the hypersurface
(for $\Omega = 1$, the unperturbed hypersurface is flat).
For adiabatic perturbations $\R$ is independent of time while outside
the horizon.

Isocurvature
perturbations\cite{EB86,MFB92,LL93} are characterized by the
entropy perturbation $ S \equiv
\delta(n_c/s_\gamma)/(n_c/s_\gamma) = \delta_c - (3/4)\delta_\gamma $, where
$n_c$ is the number density of CDM, $s_\gamma$ is the entropy
density associated with photons, and $\delta_c$ and
$\delta_\gamma$ are the relative overdensities in the CDM and
photon energy densities.  During radiation domination $S \sim
\delta_c \equiv \delta n_c/n_c$. Outside the horizon, $S$ does not
change with time.

The terms ``adiabatic" and ``isocurvature" refer to ``initial
conditions" specified at a time when the universe is
radiation-dominated and all relevant scales are well beyond the
horizon.  During this era $S$ and $\R$ are time-independent. An
adiabatic perturbation is then defined as one for which $S = 0$,
and an isocurvature perturbation as one for which $\R = 0$.  A
general perturbation is a superposition of an adiabatic and an
isocurvature perturbation (in fact, there can be several kinds of
isocurvature perturbations\cite{general} as there are several
kinds of matter; we consider here a CDM isocurvature
perturbation).

If the perturbations are Gaussian, their statistical properties
are fully described by their power spectra
$P_\R(k) \equiv \langle|\R_k|^2\rangle$ and
$P_S(k) \equiv \langle|S_k|^2\rangle$.
The spectral index is
defined by $n(k) \equiv d\ln P(k)/dk + 4$.  If the indices are
scale independent, we can write
\bea
   P_\R(k) = Ak^{\nad-4}~,
   \nonumber \\
   P_S(k) = Bk^{\niso-4}~,
\la{spectra}
\eea
where $\nad$ and $\niso$ are the spectral indices with
$\nad=1$ and $\niso = 1$ corresponding to a completely scale free spectrum.
(For $\nad$, this definition is standard,
for $\niso$, other definitions in use set the index
for a scale-free spectrum to $\niso = 0$ or $\niso = -3$.)

While the perturbations are outside the horizon, $S$ remains
time independent, but an isocurvature perturbation develops a
nonzero $\R = \frac{1}{3}S$ when the universe becomes
matter dominated\cite{LyRi99}. Thus, during last scattering a
curvature perturbation is already present. During matter
domination, the density perturbation is related to the curvature
perturbation by
\beq
   \delta_k = \frac{2}{5}\biggl(\frac{k}{aH}\biggr)^2\R_k,
\eeq
while outside the horizon ($k \ll aH$).

For low multipoles, $l \ll 200$, the corresponding scales are well
outside the horizon during last scattering and the CMB anisotropy
has a simple relation to the initial perturbations.  For adiabatic
perturbations, the anisotropy comes from the Sachs-Wolfe-effect,
\beq
   \frac{\delta T}{T} = -\frac{1}{5}\R,
\eeq
whereas for isocurvature perturbations there is also a direct
contribution from the entropy perturbation, which is 5 times
larger,
\beq
   \frac{\delta T}{T} = -\frac{1}{5}\R - \frac{1}{3}S
   = -\frac{6}{15}S = -\frac{6}{5}\R.
\eeq
(This factor 6 in $\delta T/T$ becomes a factor 36 in the $C_l$.)

For higher multipoles the perturbations have evolved by the time
of last scattering and the perturbations contribute to the
anisotropy by other mechanisms also.  The full angular power
spectra from given initial perturbations need to be calculated by
computer codes such as CMBFAST\cite{CMBFAST}.

\section{Cosmological Models}

We allow for different spectral indices (``tilts") $\nad$ and $\niso$
for adiabatic and isocurvature perturbations, and define
\beq
  \alpha \equiv \frac{C_2^{\rm iso}}{C_2^{\rm iso}+C_2^{\rm ad}}
\eeq to indicate the relative contribution of the isocurvature
fluctuations.  Here the $C_2^{\rm iso}$ and $C_2^{\rm ad}$ are the
expectation values of the isocurvature and adiabatic contributions
to the quadrupole ($C_2 = C_2^{\rm iso}+C_2^{\rm ad}$).

Note that different authors\cite{Stompor,axion,eks,gb} use
different definitions for this parameter. Since our definition
fixes the isocurvature contribution at the low end of the
multipole range, the relative isocurvature contribution at high
multipoles can be quite different depending on $\niso$ and $\nad$.
Since isocurvature perturbations produce much more power at low
multipoles, the isocurvature contribution to high multipoles is
much smaller than $\alpha$ for $\niso \approx \nad$.

Motivated by inflation, we restrict ourselves to models with
$\Omega = 1$.
We also take $r = 0$, assume $\tau = 0$ for the optical depth due to
reionization, and consider a seven-parameter space of
cosmological models. These parameters are $\Omega_\Lambda \equiv 1 -
\Omega_m$, $\omega_b \equiv \Omega_b h^2$, $\omega_c \equiv
\Omega_c h^2$, the spectral indices $\nad$ and $\niso$, and the
amplitudes $A$ and $B$ of the power spectra [Eq.~(\ref{spectra})].
The last two translate into the isocurvature contribution
parameter $\alpha$ and an overall normalization.  The Hubble
constant is given by $h =
\sqrt{(\omega_c+\omega_b)/(1-\Omega_\Lambda)}$.

In the present paper we
assume that the adiabatic and isocurvature perturbations are
uncorrelated and that their spectral indices are constant over the
cosmological range of scales. This is not necessarily the situation
in all particle physics models, but the nature of correlation
between isocurvature and adiabatic amplitudes is very much
model dependent. In general, such correlations would result in
bounds more stringent than presented here.

We use the COBE\cite{COBEdata}, Boomerang\cite{boom}, and
Maxima-1\cite{max1} data and find the best-fit models which
satisfy $h = 0.45$--0.90 and $\sigma_8\Omega_m^{0.56}$ =
0.43--0.70 (top-hat priors). We do not use any prior for the
baryon density $\omega_b$.

As noted in \cite{boom2,TZboom}, especially the
Boomerang data favors $\omega_b \sim 0.03$, which is above the
range allowed by standard big-bang nucleosynthesis, $\omega_b$ =
0.006--0.023\cite{SBBN}.  If this situation persists in the future, when
more accurate CMB data becomes available, one may need to
seriously consider non-standard nucleosynthesis scenarios\cite{NSBBN}.

We first used a coarse grid in the range
(following\cite{TegZal00}) $\Omega_\Lambda$ = 0--0.8, $\omega_b$ =
0.003--0.130, $\omega_c$ = 0.02--0.80.  We then refined it to
focus on the relevant region; for adiabatic and mixed models:
$\Omega_\Lambda$ = 0.48--0.78, $\omega_b$ = 0.018--0.042,
$\omega_c$ = 0.13--0.21, with about a dozen values for each
parameter;
 and for pure isocurvature models:
$\Omega_\Lambda$ = 0.25--0.75, $\omega_b$ = 0.004--0.028,
$\omega_c$ = 0.10--0.40. For the spectral indices we used a denser
grid with step 0.02;
in the range $\nad$ = 0.5--2.0, $\niso$ = 0.5--2.5 for adiabatic
and mixed models; and $\niso$ = 1.5--2.7 for pure isocurvature
models.

\section{Results}

The best-fit adiabatic model (model 1, see Table I) is close to
scale free, $\nad = 0.96$,
 and the fit has
$\chi^2 = 23.0$ for 30 data points and 5 parameters. (We did not
attempt to refine the parameter values between our grid points, as this is not
the focus of this paper.  See\cite{boom2,TZboom,max2} for
cosmological parameter estimation in adiabatic models for the
Boomerang and Maxima-1 data.)

The best-fit isocurvature model (model 2) has $\niso = 2.10$ and $\chi^2 =
126.4$, for 5 parameters. Clearly a pure isocurvature model is
ruled out.

\begin{table*}[t]
\begin{tabular}{lccccccccccc}
Model & $\chi^2$ & $\Omega_\Lambda$ & $
\omega_b$ & $\omega_c$ & $h$
  & $\nad$ & $\niso$ & $\alpha$ & $\alpha_{200}$ \\ \hline
(1) Best-fit adiabatic       & 23.0 & 0.60 & 0.028 & 0.19 & 0.74
  & 0.96 &&& \\
(2) Best-fit isocurvature   & 126.4 & 0.65 & 0.012 & 0.26 & 0.88
  && 2.10 && \\
(3) Best-fit mixed           & 22.5 & 0.68 & 0.030 & 0.20 & 0.85
  & 0.98 & 2.26 & 0.008 & 0.04 \\
(4) Best-fit scale-free      & 23.0 & 0.66 & 0.030 & 0.19 & 0.80
  & 1    & 1    & 0.09  & 0.002 \\
(5) Large $\alpha$ (Fig.~\ref{fig:5})     & 26.4 & 0.72 & 0.036 & 0.19 & 0.90
  & 1.20 & 0.90 & 0.63  & 0.012 \\
(6) Large $\alpha_{200}$ (Fig.~\ref{fig:6}) & 26.5 & 0.70 & 0.030 & 0.21 & 0.89
  & 1.00 & 2.06 & 0.06  & 0.13 \\
(7) Shifted, best-fit adiabatic     & 21.6 & 0.68 & 0.028 & 0.15 & 0.75
  & 0.96 &&& \\
(8) Shifted, best-fit isocurvature & 116.3 & 0.70 & 0.010 & 0.22 & 0.88
  && 2.18 && \\
(9) Shifted, best-fit mixed         & 20.9 & 0.72 & 0.030 & 0.15 & 0.80
  & 1.00 & 1.78 & 0.05 & 0.03 \\
(10) Shifted, best-fit scale-free            & 21.7 & 0.72 & 0.030 & 0.15 &
0.80 & 1 & 1    & 0.08 & 0.0016 \\
\end{tabular}
\caption[a]{\protect
Cosmological models discussed in the text and their parameter values.
``Shifted" means that the fit is done after shifting the Boomerang data up
and the Maxima-1 data down by their calibration uncertainty.
}
\end{table*}

The best fit mixed (adiabatic+isocurvature) model (model 3) has
 $\nad$ = 0.98, $\niso$ =
2.26, and $\alpha$ = 0.008.  The fit has $\chi^2 = 22.5$, for 7 parameters.
The small improvement in the fit from allowing an isocurvature
contribution indicates that the data does not suggest the presence
of such a contribution (but neither rules it out). The big blue
tilt $\niso = 2.26$ in the best-fit isocurvature contribution is
not significant, since the fit was almost as good for any $\niso$
between 0.5 and 2.5.  The isocurvature contribution is so small
that it does not affect the angular power spectrum very much, and
therefore also its tilt is not constrained. The small shifts
from changing $\niso$ can be approximated
by changing the other parameters, within the accuracy of the
present data.

\begin{figure}[tbh]
\epsfysize=7.0cm
\smallskip
\epsffile{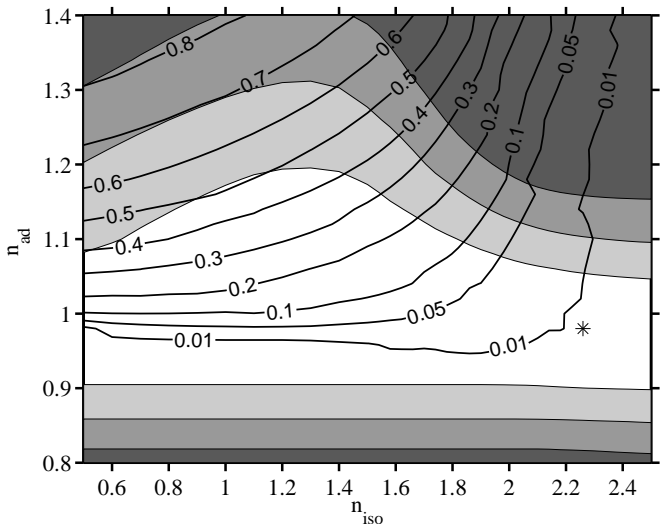}
\vspace{0.1cm}\caption[a]{\protect The 68\% (white),
95\% (light gray), and 99.7\% (medium gray) confidence level
regions ($\Delta \chi^2 = 2.3$, 6.2, and 11.8) on the
($\niso$,$\nad$)-plane, and the best-fit values of $\alpha$ for
each ($\niso$,$\nad$).  The best-fit model (model 3) is marked with
an asterisk ($\ast$).
} \label{fig:1}
\end{figure}

 For large $\niso$ the smallness of
the parameter $\alpha$  is somewhat misleading, as we defined it
as the contribution to the lowest multipole.  To represent the
isocurvature contribution in the region of the first peak, let us
define
\beq
  \alpha_{200} \equiv
  \frac{C_{200}^{\rm iso}}{C_{200}^{\rm iso}+C_{200}^{\rm ad}}~,
\eeq
where for the purpose of this paper $C_{200}$ is the average
$l(l+1)C_l$ in the 4th Boomerang bin $l$ = 176--225.  For the
above best-fit model, $\alpha_{200} = 0.04$.  Similarly, the
isocurvature contribution to $\sigma_8^2$ is 8\% in this model.
The models with smaller $\niso$ that are almost as good fits
naturally have larger $\alpha$ (see Fig.~\ref{fig:1}) but smaller
$\alpha_{200}$. Accordingly, the best-fit scale-free
($\niso$,$\nad$) = (1,1) mixed model (model 4) is almost as good a
fit, $\chi^2 = 23.0$ for only 5 parameters, but it has a larger
$\alpha$ = 0.09 and a smaller $\alpha_{200}$ = 0.002.

For the 2-$\sigma$ ($\Delta \chi^2 = 4$) upper limit to $\alpha$
we find $\alpha \leq 0.63$ and $\alpha_{200} \leq 0.13$. Within
scale-free models the upper limits are $\alpha \leq 0.34$ and
$\alpha_{200} \leq 0.010$.

In mixed models with a large $\niso$, the isocurvature contribution to
$\sigma_8$ can be significant.  The 2-$\sigma$
upper limit to the isocurvature contribution to $\sigma_8^2$ is
35\%, but within scale-free models only 0.3\%.

\begin{figure}[tbh]
\epsfysize=7.0cm
\smallskip
\epsffile{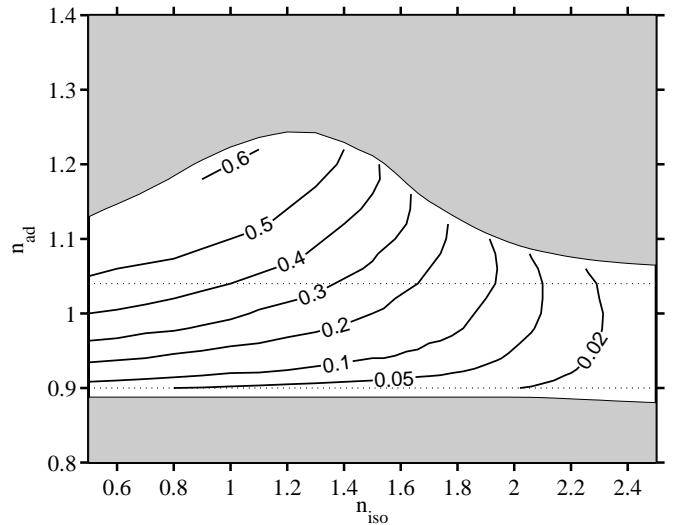}
\vspace{0.1cm}\caption[a]{\protect
The 68\%
confidence level region ($\Delta \chi^2 \leq 3.5$) in the
($\niso$,$\nad$,$\alpha$)-space represented by contours of maximum
$\alpha$. We do not show minimum $\alpha$, except that we show the
region (between the dotted lines) where the minimum is $\alpha =
0$. } \label{fig:2}
\end{figure}

In Fig.~\ref{fig:1} we show the allowed region in the
($\niso$,$\nad$)-plane, and the best-fit values of $\alpha$ for
each ($\niso$,$\nad$).
In Fig.~\ref{fig:2} we show the allowed region ($\Delta \chi^2
\leq 3.5$) in the ($\niso$,$\nad$,$\alpha$)-space as contours of
maximum $\alpha$ on the ($\niso$,$\nad$) plane.  This indicates
the magnitude of the isocurvature contribution which remains
allowed by the present data for different spectral indices.

\begin{figure}[tbh]
\vspace*{-0.3cm}
\epsfysize=7.0cm
\smallskip
\epsffile{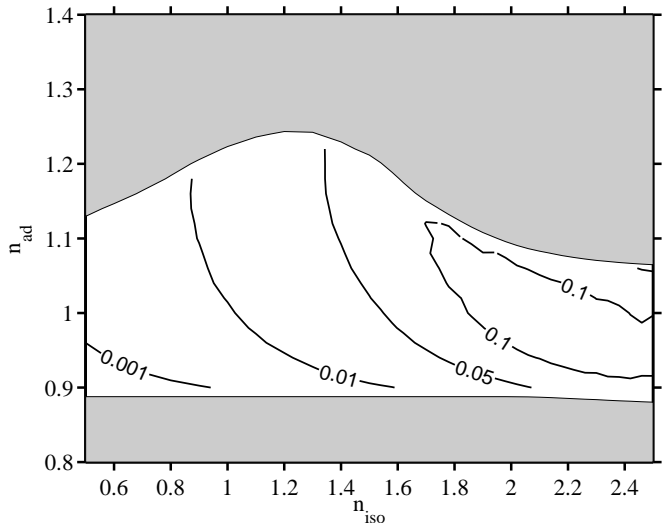}
\vspace{0.1cm}\caption[a]{\protect
Same as Fig.~\ref{fig:2}, but for $\alpha_{200}$.
}
\label{fig:3}
\end{figure}

We see that a blue tilt, $\nad > 1$, in the adiabatic spectrum
favors a larger isocurvature contribution, since the low-multipole
power from the isocurvature perturbations compensates for this
tilt.  In fact, the presence of an isocurvature contribution makes
a larger $\nad$ acceptable than in the pure adiabatic case. A
large blue tilt, $\niso \gg 1$, in the isocurvature spectrum makes
it less helpful for this purpose, and the upper limit to $\alpha$
goes down again.

Since $\alpha$ measures the contribution to the lowest multipoles,
the contribution to high multipoles for a given $\alpha$ is larger
for large $\niso$.  Thus, although the limits to $\alpha$ are
tighter for large $\niso$, this does not mean that the
contribution to high multipoles would have to be less.  In fact,
the largest allowed $\alpha_{200}$, above 0.1, are for spectral
indices $\niso > 1.8$ (see Fig.~\ref{fig:3}).

\begin{figure}[tbh]
\vspace*{-0.3cm}
\epsfysize=7.0cm
\epsffile{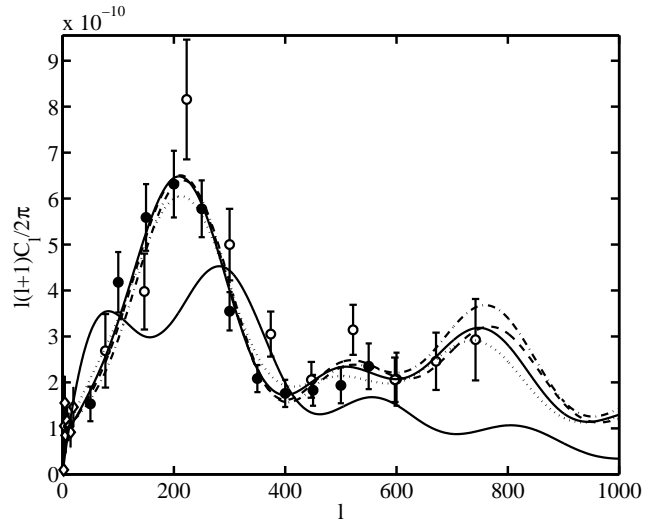}
\vspace{0.1cm}\caption[a]{\protect
The data points of COBE ($\diamond$),
Boomerang ($\bullet$), and Maxima-1 ($\circ$), and the angular
power spectra of five models: (a) Our best-fit mixed
(adiabatic+isocurvature) model (model 3, the solid line with the maximum
at $l \sim 200$). (b) The best fit adiabatic
model (model 1, dashed).
(c) The best-fit isocurvature model (model 2, the solid line with the
maximum at $l \sim 300$). (d) A mixed model with
the largest ($\Delta\chi^2 = 4$) allowed $\alpha = 0.63$
(model 5, dot-dashed). (e) A mixed model with the largest
allowed $\alpha_{200} = 0.13$ (model 6, dotted).
}
\label{fig:4}
\end{figure}

\begin{figure}[tbh]
\vspace*{-0.4cm}
\epsfysize=7.0cm
\epsffile{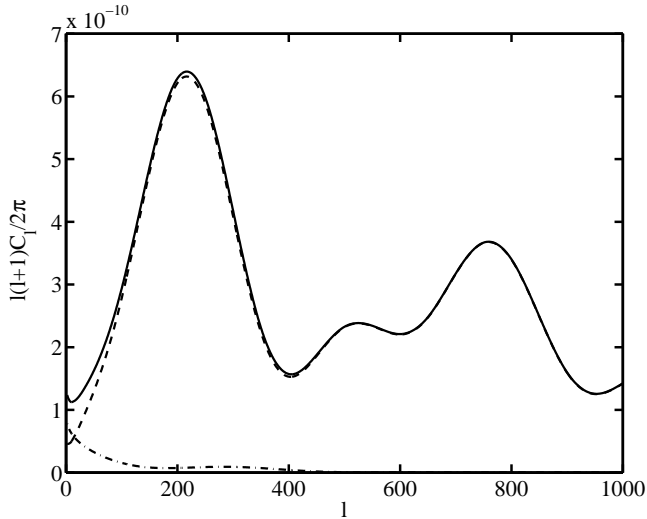}
\vspace{0.1cm}\caption[a]{\protect
The angular power spectrum, $C_l$, and the
adiabatic (dashed) and isocurvature (dot-dashed) contributions to
it, $C_l^{\rm ad}$ and $C_l^{\rm iso}$, for model 5 with $(\nad,\niso) =
(1.20,0.90)$ and $\alpha = 0.63$.
}
\label{fig:5}
\end{figure}

\begin{figure}[tbh]
\vspace*{-0.4cm}
\epsfysize=7.0cm
\epsffile{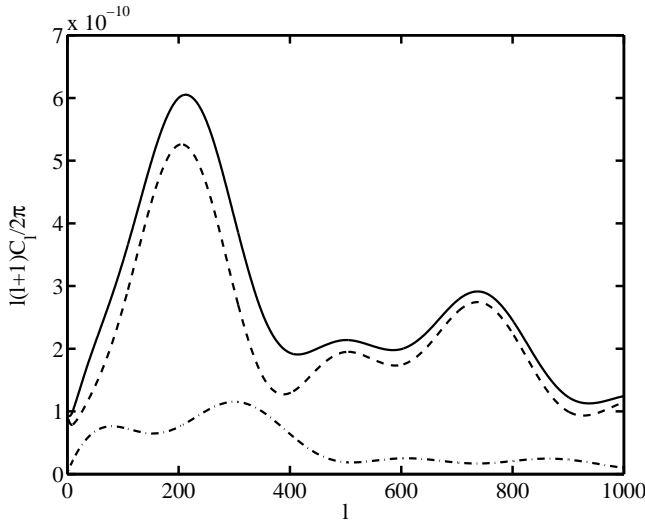}
\vspace{0.1cm}\caption[a]{\protect
Same as Fig.~\ref{fig:5}, but for model 6
with $(\nad,\niso) = (1.00,2.06)$ and $\alpha_{200} = 0.13$.
}
\label{fig:6}
\end{figure}

In Fig.~\ref{fig:4} we show the angular power spectra for some of
the models discussed above, together with the COBE, Boomerang, and
Maxima-1 data. Figures \ref{fig:5} and \ref{fig:6} show separately
the adiabatic and isocurvature contributions in models 5 and 6,
which have large isocurvature contributions (large $\alpha$ and
large $\alpha_{200}$, respectively) but are allowed by the present CMB
data.

We see how the isocurvature contribution slightly modifies the
shape of the angular power spectrum.  In model 6 it fills the
minima between the peaks in the adiabatic contribution.  An
isocurvature contribution with a large blue tilt could be the
reason for the lack of prominence of the 2nd peak in the Boomerang
data.

The Boomerang and Maxima teams estimate a calibration
uncertainty of 10\% and 4\%, respectively, in their measurements.
This uncertainty is not included in the above results.  To study
its effect, we multiplied the Boomerang measurements by $1.1^2$
and divided the Maxima-1 measurements by $1.04^2$.  This brings
the measurements closer to each other, since the Maxima-1 points
tend to lie above the Boomerang points, with the exception of the
$\l_{\rm eff} = 147$ point.  The best-fit mixed model becomes
$\nad = 1.00$, $\niso = 1.78$, and $\alpha = 0.05$ (model 9),
with an improved fit, $\chi^2 = 20.9$. The upper limit to
the isocurvature contribution is tightened to $\alpha \leq  0.56$
($\alpha_{200} \leq 0.13$ stays the same), but the results do not
change qualitatively.

The data is consistent with a purely adiabatic perturbation with
$\nad\approx 1$, with the best fit $\nad$ = 0.96  (allowing for
reionization would shift this even closer to 1). In large-field
inflation models this translates into $r \lesssim 0.3$, so that the
tensor contribution could be of the same order as the isocurvature
contribution.
Because of the
degeneracy between isocurvature and tensor perturbation such a
situation is not altogether surprising.

The COBE data together with $\sigma_8$ have an important role in
constraining the spectral indices of the models.  The Boomerang
and Maxima-1 data are important in constraining the amplitude of
the isocurvature contribution, as they outline the pattern of
acoustic peaks and thus are able to distinguish between the
different patterns of peaks\cite{HuWhite96} in the isocurvature vs
adiabatic models.

\section{Conclusions}

The main conclusion of the present work is that the Boomerang and
Maxima-1 data definitely rule out all purely isocurvature models,
including those with a large tilt \cite{Peebles1,Peebles2}.  This
is because Boomerang and Maxima-1 define the shape of the angular
power spectrum in the region of the first acoustic peaks much more
accurately than the previous data.

However, a significant isocurvature contribution remains allowed.
Indeed, as much as about half of the power at low
multipoles could come from the isocurvature contribution, although the
adiabatic contribution must dominate the first acoustic peak.
This certainly leaves much room for various particle physics
models. We should however point out that in many models, such
as two-field inflation models or Affleck-Dine models,
isocurvature and adiabatic fluctuations are not completely
independent, and much more stringent constraints on $\alpha$
(and possibly $\niso$) could be obtained in specific cases.

Note also that we cannot obtain any clear conclusion about the
spectral index $\niso$; except for large $\niso$, the correlation
between $\alpha$ and $\niso$ is small. MAP and Planck will provide
new accurate data on high multipoles and therefore could further
constrain $\niso$, but at moderate tilts we do not expect MAP to
improve much on the Boomerang and Maxima-1 limit on $\alpha$. The
situation with Planck is different, provided that high quality
polarization data is achieved as expected. Polarization will
resolve the degeneracy between isocurvature and tensor
perturbations and push the limit on $\alpha$ lower by an order of
magnitude \cite{eks}. Nevertheless, as discussed in the present
paper, already the balloon experiments on the temperature
fluctuations of the microwave sky can yield interesting
constraints on particle physics models.

\section*{Acknowledgments}

This work has been supported by the
 Academy of Finland  under the contracts 101-35224 and 101-47213.
We thank
the Center for Scientific Computing (Finland) for computational resources.
We acknowledge the use of the CMBFAST Boltzmann code developed by
Uro\v{s} Seljak and Matias Zaldarriaga.

\end{document}